\documentclass[conference,compsoc]{IEEEtran}
\usepackage{amsmath,amsfonts}
\usepackage{algorithmic}
\usepackage{algorithm}
\usepackage{array}
\usepackage[tight,footnotesize]{subfigure}
\usepackage{textcomp}
\usepackage{stfloats}
\usepackage{url}

\usepackage{verbatim}
\usepackage{graphicx}
\usepackage{pifont}
\usepackage{afterpage}
\usepackage[numbers]{natbib}
\usepackage{framed}
\usepackage{graphicx}
\graphicspath{{image/}}
\DeclareGraphicsExtensions{.pdf}
\usepackage{multirow}
\usepackage{booktabs}
\usepackage{colortbl}
\usepackage{xcolor}
\usepackage{xspace}
\usepackage{enumitem}
\usepackage{listings}
\usepackage{tabularx}
\usepackage{rotating}
\usepackage{wasysym}
\usepackage{fontawesome}
\usepackage{float}
\newcommand{\code}[1]{{\fontfamily{cmtt}\fontseries{m}\fontshape{n}\selectfont\small{#1}}}
\newcommand{\tab}{\hspace*{1em}}
\definecolor{mygray}{gray}{.9}
\newcommand{\yes}{\textcolor{green}{\ding{51}}}
\newcommand{\no}{\textcolor{red}{\ding{55}}}

\definecolor{LightGray}{gray}{0.9}

\definecolor{codegreen}{rgb}{0,0.6,0}
\definecolor{codegray}{rgb}{0.5,0.5,0.5}
\definecolor{codepurple}{rgb}{0.58,0,0.82}
\definecolor{backcolour}{rgb}{0.95,0.95,0.92}

\lstdefinelanguage{Solidity}{
	keywords=[1]{anonymous, assembly, assert, balance, break, call, callcode, case, catch, class, constant, continue, constructor, contract, debugger, default, delegatecall, delete, do, else, emit, event, experimental, export, external, false, finally, for, function, gas, if, implements, import, in, indexed, instanceof, interface, internal, is, length, library, log0, log1, log2, log3, log4, memory, modifier, new, payable, pragma, private, protected, public, pure, push, require, return, returns, revert, selfdestruct, send, solidity, storage, struct, suicide, super, switch, then, this, throw, transfer, true, try, typeof, using, value, view, while, with, addmod, ecrecover, keccak256, mulmod, ripemd160, sha256, sha3}, 
	keywordstyle=[1]\color{magenta}\bfseries,
	keywords=[2]{address, bool, byte, bytes, bytes1, bytes2, bytes3, bytes4, bytes5, bytes6, bytes7, bytes8, bytes9, bytes10, bytes11, bytes12, bytes13, bytes14, bytes15, bytes16, bytes17, bytes18, bytes19, bytes20, bytes21, bytes22, bytes23, bytes24, bytes25, bytes26, bytes27, bytes28, bytes29, bytes30, bytes31, bytes32, enum, int, int8, int16, int24, int32, int40, int48, int56, int64, int72, int80, int88, int96, int104, int112, int120, int128, int136, int144, int152, int160, int168, int176, int184, int192, int200, int208, int216, int224, int232, int240, int248, int256, mapping, string, uint, uint8, uint16, uint24, uint32, uint40, uint48, uint56, uint64, uint72, uint80, uint88, uint96, uint104, uint112, uint120, uint128, uint136, uint144, uint152, uint160, uint168, uint176, uint184, uint192, uint200, uint208, uint216, uint224, uint232, uint240, uint248, uint256, var, void, ether, finney, szabo, wei, days, hours, minutes, seconds, weeks, years},	
	keywordstyle=[2]\color{teal}\bfseries,
	keywords=[3]{block, blockhash, coinbase, difficulty, gaslimit, number, timestamp, msg, data, gas, sender, sig, value, now, tx, gasprice, origin},	
	keywordstyle=[3]\color{violet}\bfseries,
	identifierstyle=\color{black},
	sensitive=true,
	comment=[l]{//},
	morecomment=[s]{/*}{*/},
	commentstyle=\color{gray}\ttfamily,
	stringstyle=\color{red}\ttfamily,
	morestring=[b]',
	morestring=[b]"
}

\lstdefinestyle{mystyle}{
    backgroundcolor=\color{backcolour},   
    commentstyle=\color{codegreen},
    keywordstyle=\color{magenta},
    numberstyle=\tiny\color{codegray},
    stringstyle=\color{codepurple},
    basicstyle=\ttfamily\footnotesize,
    breakatwhitespace=false,         
    breaklines=true,                 
    captionpos=b,                    
    keepspaces=true,                 
    numbers=left,                    
    numbersep=5pt,                  
    showspaces=false,                
    showstringspaces=false,
    showtabs=false,                  
    tabsize=2,
    belowskip=3mm,
    aboveskip=3mm,
    escapeinside=``
}
\newcommand{\lstbg}[3][0pt]{{\fboxsep#1\colorbox{#2}{\strut #3}}}

\lstdefinelanguage{diff}{
    basicstyle=\ttfamily\small,
    morecomment=[f][\lstbg{green!20}]{+\ },
    morecomment=[f][\color{red}]{-\ },
  }

\def\BibTeX{{\rm B\kern-.05em{\sc i\kern-.025em b}\kern-.08em
    T\kern-.1667em\lower.7ex\hbox{E}\kern-.125emX}}

\begin{document}

\title{The Dark Side of Upgrades: Uncovering Security Risks in Smart Contract Upgrades}

\author{Dingding Wang, Jianting He, Siwei Wu, Yajin Zhou, Lei Wu, Cong Wang
}

\maketitle

\begin{abstract}

Smart contract upgrades are increasingly common due to their flexibility in modifying deployed contracts, such as fixing bugs or adding new functionalities. Meanwhile, upgrades compromise the immutability of contracts, introducing significant security concerns. While existing research has explored the security impacts of contract upgrades, these studies are limited in collection of upgrade behaviors and identification of insecurities.

To address these limitations, we conduct a comprehensive study on the insecurities of upgrade behaviors. First, we build a dataset containing 83,085 upgraded contracts and 20,902 upgrade chains. To our knowledge, this is the first large-scale dataset about upgrade behaviors, revealing their diversity and exposing gaps in public disclosure. Next, we develop a taxonomy of insecurities based on 37 real-world security incidents, categorizing eight types of upgrade risks and providing the first complete view of upgrade-related insecurities. Finally, we survey public awareness of these risks and existing mitigations. Our findings show that four types of security risks are overlooked by the public and lack mitigation measures. We detect these upgrade risks through a preliminary study, identifying 31,407 related issues—a finding that raises significant concerns.




\end{abstract}

\section{Introduction}
\label{sec:intro}

Smart contracts are self-executing programs deployed on blockchains. On EVM-based blockchains, smart contract code is immutable, meaning it cannot be modified once deployed. While this ensures trustworthiness, it also poses a critical challenge: vulnerabilities cannot be patched, often leading to substantial financial losses. Additionally, developers may need to introduce new features without disrupting existing user interactions. To overcome these limitations, upgradeable contracts were introduced, enabling developers to update contract logic after deployment while preserving data integrity and minimizing user disruption.

Generally, upgradeable contracts can be categorized into three primary types~\cite{frowis2022not,chen2020finding,huang2024sword,li2024characterizing,salehi2022not}: 
(i)  \code{CALL}-based upgrades, which operate by redirecting \code{CALL} instruction targets, typically adopted when decentralized applications (DApps) require module switching rather than direct code modifications within individual smart contracts;
(ii) Metamorphic upgrades, which employ opcodes \code{SELFDESTRUCT} and \code{CREATE2} to achieve complete contract replacement, though remaining uncommon in practice~\cite{li2024characterizing,huang2024sword};
and (iii) \code{DELEGATECALL}-based upgrades, the most prevalent form and de-facto industry standard~\cite{chaliasos2024smart,salehi2022not}, which separate persistent storage from upgradeable logic via \code{DELEGATECALL} to preserve state continuity across contract versions.
\code{DELEGATECALL}-based upgrades can be further classified into two implementation approaches.
The first one is \textit{proxy-based}, which is the typical and well-known form of upgrades, achieving upgradeability through a clean separation between immutable storage (proxy) and modifiable logic (implementation), connected via \code{DELEGATECALL} operations.
The second one, which we refer to as \textit{non-proxy-based}, does not rely on the conventional proxy for upgrades. Instead, these approaches employ more customized and diverse mechanisms, while still leveraging \code{DELEGATECALL} for state preservation.

\code{DELEGATECALL} facilitates flexible upgrades of individual smart contracts by permitting code modifications while maintaining persistent storage.
However, this flexibility introduces significant security considerations.
On one hand, contract logic can be modified stealthily, which enables malicious actors exploit the upgrade mechanism to inject harmful code through seemingly legitimate modifications.
On the other hand, upgrade-related vulnerabilities typically emerge in mature contracts with established user bases and substantial asset pools, unlike vulnerabilities in immutable contracts that tend to be discovered and exploited shortly after deployment.
This fact significantly amplifies potential damages when attacks occur. Our analysis of 37 upgrade-related attacks demonstrates the magnitude of these risks, with seven incidents involving more than \$10 million and two catastrophic cases surpassing \$100 million.

Security considerations about contract upgrades have attracted research efforts~\cite{bodell2023proxy,li2024characterizing,salehi2022not}. However, current research suffers from two major limitations.
First, existing work primarily focuses on proxy-based upgradeable contracts rather than upgrade behaviors themselves. This narrow focus leads to datasets that include many contracts never actually upgraded and, more critically, omits non-proxy-based upgrades entirely. For instance, in our dataset of 83,085 upgraded contracts, only 24,955 (30.0\%) comply with proxy standards (i.e., EIP-1967 and EIP-1822). Furthermore, among 1,660 proxy contracts from Ethereum Mainnet and BSC Mainnet in existing large-scale datasets~\cite{proxy_dataset}, 1,523 (91.7\%) have never performed an upgrade.
Second, existing studies lack a systematic taxonomy of security impacts rooted in empirical evidence. They predominantly study well-known issues like storage collisions, which account for only 16.2\% (6/37) of real-world incidents in our dataset. Meanwhile, high-impact risks such as malicious code injection (14 incidents, \$115M losses) and interface collisions (\$110M at risk) remain understudied. These overlooked threats underscore the urgent need for a comprehensive framework to assess upgrade-related vulnerabilities.

To address these limitations and gain a comprehensive understanding of the impact of upgrades, we conduct an in-depth study on the insecurities of upgrade behaviors. Specifically, we aim to answer the following research questions:

\begin{itemize}
    \item[\textbf{RQ1}] \textbf{Characterization of upgrade behaviors.}  
    What are the characteristics of upgrade behaviors? Are existing tools sufficient for users to accurately identify and understand these upgrades? 
    \item[\textbf{RQ2}] \textbf{Taxonomy of upgrade security.}  
    How significant is the impact of insecurities caused by upgrades? What improper aspects of upgrades are responsible for these incidents?  
    \item[\textbf{RQ3}] \textbf{Evaluation of upgrade risks in the wild.}  
    Are the security risks summarized in RQ2 publicly acknowledged? If not, how severe are their impacts in practice?  
\end{itemize}

To answer these research questions, we conduct a series of progressive studies. First, we construct a dataset of upgrade behaviors through the following steps. We begin by identifying contracts with potential upgrade behaviors, locating those that have invoked multiple \code{DELEGATECALL} targets in historical transactions. Next, we analyze the bytecode of these contracts to identify upgradeable contracts based on the sources of their \code{DELEGATECALL} targets. Finally, we detect actual upgrade behaviors by monitoring changes of their \code{DELEGATECALL} targets. Using this approach, we build a dataset containing 83,085 upgraded contracts and 20,902 upgrade chains. To our knowledge, this is the first large-scale dataset specifically focused on upgrade behaviors. During the collection process, we observe diverse patterns of upgrade behaviors. However, existing tools, which primarily consider proxy-based upgrades, fail to capture many of these upgrades, leaving users unaware of a significant portion of upgrade behaviors.

Next, we develop a comprehensive taxonomy of insecurities in upgrade behaviors based on 37 real-world security incidents involving over \$400 million in losses. The taxonomy categorizes eight types of upgrade risks into four groups: improper initialization, collisions, flawed business logic, and malicious code injection. This taxonomy provides the first complete overview of upgrade risks and establishes a foundation for further evaluation of upgrade risks.

Finally, using this taxonomy, we conduct a comprehensive survey to assess public awareness of these risks and the availability of mitigation measures. Our survey reveals that the security impact of five out of the eight risk types remains unknown to the community, and four of these are entirely overlooked by the public, lacking any mitigation measures. We further perform a preliminary study to detect three types of risks—improper initialization, collisions, and suspicious code injection—in our dataset. In total, we identify 31,407 related issues, highlighting a significant and alarming concern.


In summary, our work makes the following contributions.  
\begin{itemize}[leftmargin=*]  
    \setlength{\itemsep}{0pt}  
    \setlength{\parsep}{0pt}  
    \setlength{\parskip}{0pt}  
    \item We build the first large-scale dataset\footnote{All study data/artifacts will be open-sourced to facilitate future research} on upgrade behaviors, shedding light on the diversity of upgrade behaviors and exposing the limitations of existing tools.
    \item We develop a comprehensive taxonomy of the security risks associated with upgrade behaviors based on 37 real-world security incidents involving more than \$400 million. The taxonomy includes eight different types of insecurities, with half of them being publicly overlooked.  
    \item Following the insecurity taxonomy, we performed a preliminary study to detect upgrade risks in the wild and identified a total of 31,407 issues.  
\end{itemize}

\section{Background}
\label{Sec:Background}

\subsection{Smart Contracts}
\label{subsec:EVM}

Ethereum~\cite{wood2014ethereum} and Binance Smart Chain (BSC)~\cite{bscscan} are two leading programmable blockchains supporting smart contracts. These contracts often handle financial functionalities involving native tokens (e.g., ETH) and standardized tokens like ERC-20~\cite{erc20} and ERC-721~\cite{erc721}, which define interfaces such as \code{transfer} and \code{transferFrom} for token transactions.

Smart contracts operate using two account types: Externally Owned Accounts (EOAs) and contract accounts. EOAs, controlled by private keys, initiate transactions. Contract accounts store publicly accessible bytecode. When an EOA initiates an external transaction invoking a contract account, internal transactions may occur if the contract invokes another via instructions like \code{DELEGATECALL}. The external transaction initiator (the EOA) is \code{tx.origin}, while the internal initiator (the caller contract) is \code{msg.sender}.

Contract accounts use storage for persistent variables. Storage is a 32-byte slot array indexed from 0 to $2^{256} - 1$, employing a packed layout to combine smaller items into single slots. The \code{SLOAD} instruction reads full 32-byte slots, with bit-masking used to access packed variables.

\subsection{DELEGATECALL-based Upgradeable Smart Contracts}

\begin{figure}[t]
    \centering
    \includegraphics[width = 0.8\linewidth]{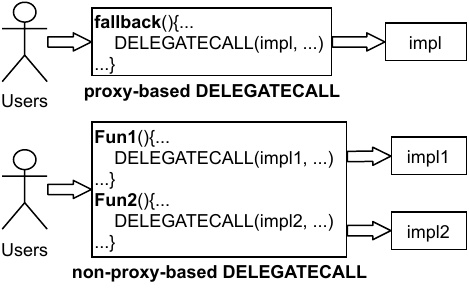}
    \vspace{-0.3cm}
    \caption{Different types of \code{DELEGATECALL}-based upgradeable contracts.}
    \label{fig:delegatecall_based}
    \vspace{-0.4cm}
\end{figure}

The \code{DELEGATECALL} operation enables a contract to execute code from other contracts while maintaining its own execution context such as storage. This characteristic facilitates flexible code composition by embedding logic from other contracts, and enables upgradeability by simply modifying the \code{DELEGATECALL} target in storage to point to new implementations.
Based on implementation approaches, \code{DELEGATECALL}-based upgradeable contracts can be classified into two categories: proxy-based and non-proxy-based architectures. As illustrated in Figure~\ref{fig:delegatecall_based}.

Proxy-based contracts consist of a state-holding proxy and a code-holding implementation contract.
The proxy does not contain business logic, but serving as a persistent entry point that maintains all storage variables while delegating code execution to the implementation.
The upgrades of proxy-based contracts are executed by replacing the \code{impl} contract that contains all business logic.
Standards like EIP-1967 and EIP-1822 formalize this design. EIP-1967 specifies slots such as \code{0x36..bc}\footnote{(bytes32(uint256(keccak256('eip1967.proxy.implementation')) - 1), \\0x360894a13ba1a3210667c828492db98dca3e2076cc3735a920a3ca505d382bbc} for the implementation address and \code{0xb5..03}\footnote{(bytes32(uint256(keccak256('eip1967.proxy.admin')) - 1), \\0xb53127684a568b3173ae13b9f8a6016e243e63b6e8ee1178d6a717850b5d6103} for the admin. 
EIP-1822 introduces the Universal Upgradeable Proxy Standard (UUPS), embedding upgrade logic into the implementation and using slot \code{0xc5..f7}\footnote{(keccak256("PROXIABLE")),\\ 0xc5f16f0fcc639fa48a6947836d9850f504798523bf8c9a3a87d5876cf622bcf7} for the implementation address.

While proxy-based upgradeable contracts have received considerable attention from both researchers and developers, non-proxy-based ones remain largely overlooked. Unlike standardized patterns in proxies, non-proxy ones exhibit greater diversity without unified conventions, as developers can freely customize the way they incorporate external contract logic through \code{DELEGATECALL}.
Figure~\ref{fig:delegatecall_based} illustrates one type of them.\footnote{For exmaple, 0x6e16394cbf840fc599fa3d9e5d1e90949c32a4f5 on Ethereum.} In this case, functions \code{Fun1} and \code{Fun2} extend functionality with custom additions based on the logic from \code{impl1} and \code{impl2}, respectively.
This approach enables more granular upgrades, allowing developers to update specific functionalities by replacing individual implementation contracts, for example, modifying \code{Fun1} by replacing \code{impl1}.

To generalize our study, we use “caller contract” for the \code{DELEGATECALL} initiator and “callee contract” for the recipient, replacing the proxy/implementation terminology. This ensures broader applicability and accuracy in describing \code{DELEGATECALL}-based upgrades, encompassing both proxy-based and non-proxy designs.



\section{Overview}
\label{sub:overview}
The purpose of this study is to provide a comprehensive understanding of smart contract upgrades. Before delving into the details of this work, we define the key terms used throughout the paper.

\begin{itemize}[leftmargin=*]
    \item \textbf{Upgrade Mechanism} This paper focuses on the upgrade of deployed code based on changing the target of the instruction \code{DELEGATECALL}. The smart contract that uses the \code{DELEGATECALL} is called the \textit{caller contract}, while the target of the \code{DELEGATECALL} is referred to as the \textit{callee contract}.
    \item  \textbf{Upgradeable Smart Contracts} Smart contracts that contain \code{DELEGATECALL}, with the target stored in the storage. 
    \item \textbf{Upgrade Behavior} The fact that a upgradeable contract changes the target of the \code{DELEGATECALL}.
    \item \textbf{Upgraded Smart Contracts} Upgradeable contracts that have performed the upgrade behavior.   
    \item \textbf{Upgrade Chain}  A sequence of callee contracts used by a caller contract across multiple upgrades. For example, if caller contract A upgrades by switching from callee B to C to D, its upgrade chain is B$\rightarrow$C$\rightarrow$D.
\end{itemize}





\begin{figure}[t]
    \centering
    \includegraphics[width = 0.97\linewidth]{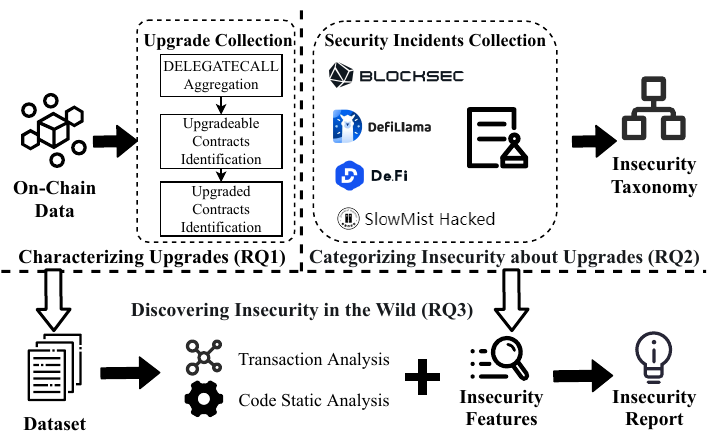}
    \vspace{-0.3cm}
    \caption{Overview of Our Study.}
    \label{fig:overview}
    \vspace{-0.4cm}
\end{figure}

Fig.~\ref{fig:overview} shows the overall process of our study.
First, we combine transaction analysis and code analysis to collect upgrade behaviors based on \code{DELEGATECALL} in Ethereum and BSC, and collect 83,085 upgraded contracts with 20,902 upgrade chains (\S\ref{sec:RQ1}). To our knowledge, it is the first large-scale dataset about upgrade behaviors, revealing the diversity of upgrade methods.
Second, we collect reported security incidents about upgrades from four authority institutions in the community, and analyze their root causes. We collect 37 incidents involving more than \$400 million, and classify their root causes into four categories (\S\ref{sec:RQ2}). To our knowledge, we are the first to conduct a comprehensive study on security incidents about upgrades and develop an insecurity taxonomy.
Third, we extract transaction-related and code-related symptoms about insecurities found in RQ2, and discover a total of 31,407 real-world security issues in three publicly unknown categories based on the dataset constructed in RQ1 (\S\ref{sec:RQ3}).

\section{Characterization of Upgrades (RQ1)}
\label{sec:RQ1}

To answer RQ1, we construct the first large-scale dataset about upgrade behaviors.
We combine transaction analysis and bytecode analysis to detect upgrades in two popular programmable blockchains, Ethereum and BSC.
After collecting upgrades, we characterize them from upgrade methods and their upgrade activity.
Finally, we reveal issues we find during collection, including erroneous upgrade targets and opacity upgrade process.

\subsection{Dataset Construction}

We construct a dataset of upgrade behaviors by analyzing \code{DELEGATECALL} usage in Ethereum and BSC transactions. Our methodology involves three steps: (1) identifying \code{DELEGATECALL} invocations through transaction traces, (2) filtering upgradeable contracts by analyzing their \code{DELEGATECALL} targets, and (3) detecting upgraded contracts by monitoring changes of \code{DELEGATECALL} targets of upgradeable contracts. This approach ensures we capture genuine upgrade behavior.

\noindent\textbf{DELEGATECALL Aggregation}\tab
This step analyzes historical transaction traces to identify contracts that have invoked \code{DELEGATECALL} to call multiple different targets. By focusing on actual invocations, we exclude contracts that contain \code{DELEGATECALL} in their bytecode but never invoke it or only call a single contract, as these do not exhibit upgrade behavior. This significantly reduces the analysis scope. Specifically, for each transaction, we extract execution traces, record the caller and callee contract addresses of \code{DELEGATECALL}, and retain only those caller contracts that interact with multiple callee contracts as potentially upgradeable.

\noindent\textbf{Upgradeable Contracts Identification}\tab
We refine the dataset by identifying truly upgradeable contracts, as not all contracts invoking \code{DELEGATECALL} are upgradeable. For instance, if the \code{DELEGATECALL} target is hardcoded in the bytecode, it cannot be modified, rendering the contract non-upgradeable. Similarly, if the target is passed through calldata, the contract typically acts as a forwarder without consistent functionality, as the target depends on external input and is not controlled by the caller. Therefore, a contract is considered upgradeable only if its \code{DELEGATECALL} target is both mutable and controllable.

 To identify such contracts, we perform static analysis on the bytecode to trace the provenance of \code{DELEGATECALL} targets. Our methodology specifically examines the second operand of each \code{DELEGATECALL} instruction (representing the target address) and recursively traces its data dependencies through def-use chain analysis. We only classify a contract as upgradeable when our analysis confirms that the \code{DELEGATECALL} target ultimately derives from storage, which indicates a mutable and controllable execution target.




\noindent\textbf{Upgraded Contracts Identification}\tab
After identifying upgradeable contracts, we detect upgrade behaviors by monitoring changes in the \code{DELEGATECALL} targets. Specifically, upgrade behaviors occur when \code{DELEGATECALL} targets change. To locate the target in storage, we use a \code{(slot index, offset)} pair. The slot index, retrieved from the operand of the \code{SLOAD} instruction, identifies a 32-byte storage slot, while the offset represents the position of the 20-byte address within the slot, as addresses can be packed with other variables. By tracking historical transactions, we record changes in these storage slots and extract the sequence of target addresses. These addresses are then used to construct the \textit{upgrade chain} for each upgraded contract, representing all their historical callee contracts.

\subsection{Upgrade Behavior Analysis}

We collect upgrades in two representative EVM-compatible blockchains, Ethereum (first 18,000,000 blocks) and BSC (first 32,000,000 blocks), constructing a dataset of $830,387$ upgradeable contracts, $83,085$ upgraded contracts, and $20,902$ upgrade chains.
Only about 10\% of upgradeable contracts undergo upgrades, indicating a relatively low upgrade rate. Notably, the significant difference between the number of upgraded contracts and upgrade chains arises because multiple caller contracts often share the same upgrade chain, e.g., $65,196$ upgraded contracts on Ethereum share only $7,123$ upgrade chains. This is common in projects like wallets, where many caller contracts share the same callee contracts, amplifying risks as a single vulnerability in a callee contract could impact thousands of associated caller contracts.

Contrary to the common belief that upgrades are predominantly proxy-based, we observed significant diversity in upgrade methods, reflected in 196 distinct storage locations for \code{DELEGATECALL} targets, classified into four types:

\begin{table}
\centering
\caption{Slot Usage Distribution, differing between Ethereum and BSC.}
\vspace{-0.3cm}
\footnotesize
\label{tab:slot_usage}
\resizebox{\linewidth}{!}{%
\begin{tabular}{llccccccc}
\toprule
\multirow{2}{*}{} & \multirow{2}{*}{} & \multirow{2}{*}{\begin{tabular}[c]{@{}c@{}}Simple\\Slot Key\end{tabular}} & \multicolumn{3}{c}{Complex Slot Key} & \multirow{2}{*}{\begin{tabular}[c]{@{}c@{}}Multiple\\Slots\end{tabular}} & \multirow{2}{*}{\begin{tabular}[c]{@{}c@{}}Slots with\\Offset\end{tabular}} \\
\cmidrule(lr){4-6}
 &  &  & EIP-1967 & EIP-1822 & others &  &  &  \\
\midrule
& upgradeable & \textbf{772,106} & 16,808 & 55 & 1,245 & 76 & 12,614 \\
& upgraded & 55,579 & 9,108 & 51 & 438 & 30 & 10,744 \\
& upgrade chains & 750 & 5,798 & 40 & 312 & 30 & 3 \\ 
\multirow{-4}{*}{Ethereum}  & upgrade rate & 7.20\% & 54.62\% & 92.73\% & 35.18\% & 39.47\% & 85.18\% \\
\midrule
& upgradeable & 13,748 & \textbf{25,249} & 63 & 1,124 & 678 & 16 \\
& upgraded & 1,867 & 15,737 & 59 & 221 & 27 & 15 \\
& upgrade chains & 829 & 12,481 & 57 & 199 & 21 & 15 \\
\multirow{-4}{*}{BSC} & upgrade rate & 13.58\% & 62.33\% & 93.65\% & 19.66\% & 3.98\% & 93.75\% \\
\bottomrule
\end{tabular}
}
\end{table}

\setlength{\itemindent}{-15pt}
\setlength{\labelsep}{5pt}
\begin{itemize}[leftmargin=*]
\setlength{\itemsep}{0pt}
\setlength{\parsep}{0pt}
\setlength{\parskip}{0pt}
\item \textbf{Simple slot key}, e.g., 0x0, prone to collisions.
\item \textbf{Complex slot key}, e.g., 0x69ed75 or slots specified by EIP-1967/EIP-1822, often computed using hash functions or specified by developers.
\item \textbf{Multiple slots}, used when caller contracts employ multiple callee contracts simultaneously.
\item \textbf{Slots with offset}, where callee addresses are packed in storage with other variables.
\end{itemize}

Table~\ref{tab:slot_usage} shows the slot usage distribution. Contracts adhering to EIP-1967 and EIP-1822 demonstrate higher levels of upgrade activity, particularly EIP-1822 contracts.  Some contracts adhering to these standards undergo nearly a hundred upgrades, while the majority are upgraded fewer than ten times.


\begin{table}
\centering
\caption{Issues in Upgrade}
\vspace{-0.3cm}
\label{tab:upgrade_issues}
\footnotesize
\resizebox{\linewidth}{!}{%
\begin{tabular}{cccc} 
\toprule
 & Redundant Upgrade & Invalid Upgrade & Upgrade to Zero Addr. \\ 
\midrule
Ethereum & 497 & 59 & 29 \\
BSC & 1,370 & 38 & 18 \\
\midrule
Total & 1,867 & 97 & 47 \\
\bottomrule
\end{tabular}
}
\end{table}

\begin{table}[t]
\centering
\caption{Issues about Existing Blockchain Explorers. USC is short for upgraded smart contract, and UC is short for upgrade chains.}
\vspace{-0.3cm}
\footnotesize
\label{tab:explorer_issues}
\resizebox{\linewidth}{!}{%
\begin{tabular}{lccccc} 
\toprule
 & \multicolumn{3}{c}{\begin{tabular}[c]{@{}c@{}}Blockchain Explorers\\(Etherscan/Bscscan)\end{tabular}} & \multicolumn{2}{c}{\begin{tabular}[c]{@{}c@{}}Upgrade Explorer\\(Upgradehub)\end{tabular}} \\ 
\cmidrule(lr){2-4}\cmidrule(lr){5-6}
 & \begin{tabular}[c]{@{}c@{}}Unrecognized\\USC\end{tabular} & \begin{tabular}[c]{@{}c@{}}Incomplete\\Logic\end{tabular} & \begin{tabular}[c]{@{}c@{}}Fake\\Logic\end{tabular} & No UC & \begin{tabular}[c]{@{}c@{}}Incomplete\\UC\end{tabular} \\ 
\midrule
Ethereum & 53,091 & 61 & 319 & 63,307 & 813 \\
BSC & 11,182 & 663 & 0 & 15,337 & 867 \\ 
\midrule
Total & 64,273 & 724 & 319 & 78,644 & 1,680 \\
\bottomrule
\end{tabular}
}
\end{table}

\begin{figure}[t]
\centering
\footnotesize
\begin{lstlisting}[language=Solidity]
contract SolidlyProxy{
  bytes32 constant IMPLEMENTATION_SLOT = 0x36..bc; 
// keccak256('eip1967.proxy.implementation'), actually used for interface so etherscan picks up the interface
  bytes32 constant LOGIC_SLOT = 0x59..ab;
// keccak256('LOGIC')-1, actual logic implementation
...}
\end{lstlisting}
\vspace{-0.3cm}
\caption{The code snippet that causes Etherscan to fail to recognize the correct implementation address.}
\label{lst:slot_cheat}
\end{figure}

\subsection{Observed Security Issues}  
In the following, we elaborate on the observed security issues in contract upgrades, including erroneous upgrade targets and opacity in the upgrade process. Erroneous upgrade targets refer to inappropriate new callee addresses during upgrades, a phenomenon reported by existing works~\cite{bodell2023proxy,li2024characterizing}. Table~\ref{tab:upgrade_issues} outlines the issues we identified: redundant upgrades (reusing previous callee addresses), invalid upgrades (non-executable addresses like EOAs), and upgrades to a zero address. Redundant upgrades waste resources and may reintroduce vulnerabilities, while invalid upgrades and upgrades to a zero address render contracts non-functional, highlighting chaotic upgrade management.

More importantly, we discovered significant challenges in observing upgrade behaviors, making the upgrade process opaque. Although blockchain explorers like Etherscan~\cite{etherscan} and BscScan~\cite{bscscan} label upgradeable contracts and their \code{DELEGATECALL} callees, and tools like Upgradehub~\cite{upgradehub} exist to track upgrades, their information is often incomplete. These tools primarily recognize one-to-one proxy-to-logic upgrade patterns, failing to detect unconventional methods or multiple-slot upgrades. As a result, many upgraded contracts remain unrecognized, and complete logic information is often unavailable. The issues with these explorers regarding upgrade behavior disclosure are detailed in Table~\ref{tab:explorer_issues}.

Worse, these explorers can be intentionally deceived to display incorrect logic, misleading users. For instance, Fig.~\ref{lst:slot_cheat} shows a contract\footnote{e.g., 0x77779759974f2353835F1A8c17B88f6F1f3e4362 on Ethereum} using two slots to store callees: \code{0x36..bc} (EIP-1967) and \code{0x59..ab} (real implementation). Since explorers prioritize EIP-specified slots, only the fake callee stored in \code{0x36..bc} is displayed, concealing the actual callee address and preventing users from knowing the true code executed. This deception undermines transparency and poses severe security risks, as users interact with contracts under false assumptions about their functionality. Such opacity can lead to exploitation, as malicious actors can hide vulnerabilities or malicious logic behind misleading explorer displays, leaving users—and even security analysts—unaware of the true risks.

\begin{framed}
    \vspace{-0.1in}
        \noindent
        \textbf{Answer to RQ1}\tab
        Upgrade behaviors are diverse, hindering existing tools from providing accurate and transparent information about upgrades.
        \vspace{-0.1in}
\end{framed}


\begin{table*}[t]

\newcommand{\initTable}{
    \begin{tabularx}{435pt}{XXXll}
    \textbf{Incident} & \textbf{Date} & \textbf{Chain} & \textbf{Loss} & \textbf{Root Cause} \\
    Parity & 2017-07-19 & Ethereum & \textasciitilde{}\$34M & public initialization  \\
    Parity & 2017-11-08 & Ethereum & \textasciitilde{}\$155M & uninitialized \\
    L2DAO & 2022-10-22 & Optimism & 49.95M L2DAO tokens & failed initialization \\
    Ronin Bridge & 2024-08-06 & Ethereum & \textasciitilde{}\$12M & incomplete initialization \\
    \end{tabularx}
}

\newcommand{\collision}{
\begin{tabularx}{435pt}{lXlll}
    \textbf{Incident} & \textbf{Date} & \textbf{Chain} & \textbf{Loss} & \textbf{Root Cause} \\
    EFVault & 2023-02-24 & Ethereum & \textasciitilde{}\$5.1M & callee/callee storage collision  \\
    AAVE & 2023-05-20 & Polygon & freezing  \textasciitilde{}\$110M & interface collision  \\
    TelCoin & 2023-12-26 & Polygon & \textasciitilde{}\$1.24M & caller/callee storage collision  \\
    Ember Sword & 2024-04-27 & Polygon & \textasciitilde{}\$196k & caller/callee storage collision  \\
    PikeFinance & 2024-04-30 & Ethereum, Arbitrum, Optimism & \textasciitilde{}\$1.7M & callee/callee storage collision  \\
    OKX NFT Aggregator & 2024-06-20 & BSC & \textasciitilde{}\$14k & callee/callee storage collision  \\
    DeltaPrime & 2024-07-22 & Arbitrum & \textasciitilde{}\$1M & caller/callee storage collision  \\
\end{tabularx}
}

\newcommand{\logic}{
\begin{tabularx}{435pt}{lXlXlcc}
    \textbf{Incident} & \textbf{Date} & \textbf{Chain} & \textbf{Loss} & \textbf{Vulnerability Type} & \textbf{Fix} & \textbf{Introduce}  \\
    Ankr & 2022-12-02 & BSC & \textasciitilde{}\$5M & public mint & \yes & \ding{109}  \\
    Thoreum Finance & 2023-01-19 & BSC & \textasciitilde{}\$580k & self transfer & \yes & \ding{109} \\
    BSCANT3 & 2023-01-19 & BSC & 1466 BNB & public burn & \yes & \RIGHTcircle  \\
    Indexed Finance & 2023-03-20 & Ethereum & \textasciitilde{}\$12k & self transfer & \yes & \ding{109}  \\
    SafeMoon & 2023-03-28 & BSC & \textasciitilde{}\$8M & public mint, public burn & \yes & \RIGHTcircle \ding{109}  \\
    MetaPoint & 2023-04-11 & BSC & \textasciitilde{}\$920k & public approve & \yes & \ding{109} \\
    Level Finance & 2023-05-01 & BSC & \textasciitilde{}\$1M & inconsistent accounting & \yes & \RIGHTcircle \\
    DEI & 2023-05-05 & Ethereum, BSC, Arbitrum & \textasciitilde{}\$6M & accounting error & \no & \RIGHTcircle \\
    Floor Protocol & 2023-12-17 & Ethereum & \textasciitilde{}\$1.6M & arbitrary call & \yes & \RIGHTcircle \\
    Socket & 2024-01-16 & Ethereum & \textasciitilde{}\$3.2M & unverified user input & \yes & \RIGHTcircle \\
    XBridge & 2024-04-24 & Ethereum, BSC & \textasciitilde{}\$1.8M & unverified user input & \no & \ding{109} \\
    Bedrock uniBTC & 2024-09-26 & Ethereum & \textasciitilde{}\$2M & state validation issue & \yes & \ding{109} \\
\end{tabularx}
}

\newcommand{\malicious}{
\begin{tabularx}{435pt}{lllXlclc}
    \textbf{Incident} & \textbf{Date} & \textbf{Chain} & \textbf{Loss} & \textbf{Root Cause} & \textbf{Introduce} & \textbf{Victim Type} & \textbf{Code Features} \\
    Meerkat Finance & 2021-03-04 & BSC & \textasciitilde{}\$32M & Rug Pull & \ding{108} & vault & \textbf{C(EB)} + \textbf{P(D)} \\
    Bent Finance & 2021-12-21 & Ethereum & \textasciitilde{}\$1.75M & \faKey & \RIGHTcircle & pool & \textbf{C(EB)} + \textbf{P(D)} \\
    OCASH & 2022-11-29 & BSC & \textasciitilde{}\$30k & Rug pull & \RIGHTcircle & sale & \textbf{C(EB)} + \textbf{P(D)} \\
    ARV Token & 2022-12-02 & BSC & \textasciitilde{}\$506k & \faKey & \RIGHTcircle & token & \textbf{C(EB)} + \textbf{P(D)} \\
    Atlantis Loans & 2023-06-10 & BSC & \textasciitilde{}\$1M & vote & \ding{108} & token & \textbf{C(EB)} + \textbf{P(D)} \\
    OKX DEX & 2023-12-13 & Ethereum & \textasciitilde{}\$2.7M & \textcolor{orange}{\faKey} & \ding{108} & privilege & \textbf{C(E)} + \textbf{P(I)} \\
    Concentric Finance & 2024-01-22 & Arbitrum & \textasciitilde{}\$1.7M & \textcolor{orange}{\faKey} & \RIGHTcircle & vault & \textbf{C(E)} + \textbf{P(D)} \\
    Shido & 2024-02-29 & Ethereum & \textasciitilde{}\$3.3M & \faKey & \RIGHTcircle & staking & \textbf{C(E)} + \textbf{P(D)} \\
    Polyhedra & 2024-03-12 & BSC & \textasciitilde{}\$760k & \textcolor{orange}{\faKey} & \ding{108} & bridge & \textbf{C(B)} + \textbf{P(D)} \\
    Wilder World & 2024-03-16 & Ethereum & \textasciitilde{}\$1.8M & \textcolor{orange}{\faKey} & \RIGHTcircle & vesting & \textbf{C(B)} + \textbf{P(D)} \\
    Curio Ecosystem & 2024-03-23 & Ethereum & \textasciitilde{}\$16M & vote & \ding{108} & privilege & \textbf{P(I)} \\
    Perpy Finance & 2024-05-06 & Arbitrum & \textasciitilde{}\$132k & public upgrade & \ding{108} & staking & \textbf{C(E)} + \textbf{P(D)} \\
    OpSec & 2024-07-11 & Ethereum & \textasciitilde{}\$182k & \faKey \textcolor{orange}{\faKey} & \ding{109} & staking & \textbf{C(E)} + \textbf{P(D)} \\
    Radiant & 2024-10-16 & BSC, Arbitrum & \textasciitilde{}\$58M & \faKey \faKey & \ding{108} & lending pool & \textbf{C(E)} + \textbf{P(D)} \\
\end{tabularx}
}

\centering
\caption{Incidents about upgrade. \ding{108} represents replacing origin contract to a malicious contract. \ding{109} represents changing existing function and make it malicious or vulnerable. \RIGHTcircle represents adding malicious or vulnerable functions into the origin contract. \faKey represents private key compromised of the owner of the contract and \textcolor{orange}{\faKey} represents private key compromised of owner of privileged contract. In code features, \textbf{C} represents constraints, which can on executor (\textbf{E}) or beneficiary (\textbf{B}). \textbf{P} represents profit, which can be direct (\textbf{D}) or indirect (\textbf{I}).}
\label{tab:incidents}
\footnotesize
\begin{tabular}{|c|c|c|@{}l|} 
\hline
\multirow{5}{*}{\rotatebox{90}{Transaction}} & \multicolumn{2}{c|}{\multirow{5}{*}{\rotatebox{90}{
\begin{minipage}{1.4cm}
Improper \\
Initialization
\end{minipage}
}}} & \multirow{5}{*}{\initTable} \\
 & \multicolumn{2}{c|}{} & \\
  & \multicolumn{2}{c|}{} & \\
 & \multicolumn{2}{c|}{} & \\
 & \multicolumn{2}{c|}{} & \\
\hline
\multirow{36}{*}{\rotatebox{90}{Code}} & \multirow{21}{*}{\rotatebox{90}{Vulnerable}} & \multirow{8}{*}{\rotatebox{90}{Collision}} & \multirow{8}{*}{\collision} \\
 & & & \\
 & & & \\
 & & & \\
 & & & \\
 & & & \\
 & & & \\
 & & & \\
\cline{3-4}
 &  & \multirow{13}{*}{\rotatebox{90}{Flawed Business Logic}} & \multirow{13}{*}{\logic} \\
 & & & \\
 & & & \\
 & & & \\
 & & & \\
 & & & \\
 & & & \\
 & & & \\
 & & & \\
 & & & \\
 & & & \\
 & & & \\
 & & & \\
\cline{2-4}
 & \multicolumn{2}{c|}{\multirow{15}{*}{\rotatebox{90}{Malicious Code Injection}}} & \multirow{15}{*}{\malicious} \\
 & \multicolumn{2}{c|}{} & \\
 & \multicolumn{2}{c|}{} & \\
 & \multicolumn{2}{c|}{} & \\
 & \multicolumn{2}{c|}{} & \\
 & \multicolumn{2}{c|}{} & \\
 & \multicolumn{2}{c|}{} & \\
 & \multicolumn{2}{c|}{} & \\
 & \multicolumn{2}{c|}{} & \\
 & \multicolumn{2}{c|}{} & \\
 & \multicolumn{2}{c|}{} & \\
 & \multicolumn{2}{c|}{} & \\
 & \multicolumn{2}{c|}{} & \\
 & \multicolumn{2}{c|}{} & \\
 & \multicolumn{2}{c|}{} & \\
\hline
\end{tabular}

\end{table*}

\section{Taxonomy of Upgrade Risks (RQ2)}
\label{sec:RQ2}

To answer RQ2, we develop a taxonomy of insecurity in upgrade behaviors based on real-world security incidents.
These incidents are collected from four well-known security incident databases in the community: Blocksec Security Incidents~\cite{blocksec}, DefiLlama Hacks~\cite{defiLlama}, De.Fi Rekt Database~\cite{de.fi}, and SlowMist Hacked~\cite{slowmist}.  
Overall, we document 37 incidents involving over \$400 million, and classified them based on their root causes, as shown in Table~\ref{tab:incidents}.  

\begin{figure}[t]
    \centering
    \includegraphics[width = 0.9\linewidth]{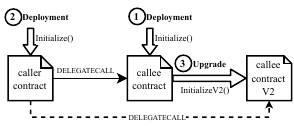}
    \vspace{-0.3cm}
    \caption{Lifecycle of upgradeable smart contracts. $\Rightarrow$ represents transactions about upgradeable contracts and $\rightarrow$ represents invocations between contracts.}
    \label{fig:init_process}
    \vspace{-0.4cm}
\end{figure}

Specifically, there are two main fundamental classes: transaction-related and code-related.  
Transaction-related causes refer to misbehavior within transactions (\ding{172}\ding{173}\ding{174} in Fig.~\ref{fig:init_process}) that support the upgrade mechanism.  
Code-related causes point to inappropriate or anomalous occurrences within the code of contracts associated with upgrades (caller/callee/callee V2 in Fig.~\ref{fig:init_process}).  

On this basis, transaction-related incidents are primarily attributed to improper initialization (\S\ref{subsec:init}), while code-related incidents can be divided into vulnerable incidents and malicious incidents (\S\ref{subsec:malicious}).  
Vulnerable incidents can be further categorized into collisions (\S\ref{subsec:collision}) and business logic flaws (\S\ref{subsec:logic}).  
The insecurity in each subclass is detailed in the following sections.

\subsection{Improper Initialization}
\label{subsec:init}

To support the upgrade mechanism, upgradeable contracts involve three types of initialization, as shown in Fig.~\ref{fig:init_process} (transactions \ding{172}, \ding{173}, and \ding{174}): (1) initialization of callee contracts, (2) initialization of the caller contract, and (3) initialization of variables introduced by upgrades. The first two typically occur during deployment, while the third occurs during upgrades. These initializations are unique to upgradeable contracts because their setup cannot rely on traditional \code{Constructor}; instead, they require special \code{initialize} functions to ensure proper initialization across upgrades. Generally, the \code{initialize()} function is public, and the caller of this function becomes the owner of the contract, making proper initialization critical to prevent unauthorized access.

\noindent\textbf{Type I. Initialization of callee contracts (\ding{172}).}\tab  
Although callee contracts typically do not hold storage, the Parity incident, which resulted in a \$155 million loss, highlights the critical importance of properly initializing them. This incident occurred because attackers exploited uninitialized callee contracts, gaining ownership and using privileged functions to destroy their code via the \code{SELFDESTRUCT} instruction, rendering the funds locked and unusable forever. In the EVM, \code{SELFDESTRUCT} can permanently destroy a contract’s code. Even if the callee contract does not contain a \code{SELFDESTRUCT} instruction, once it features \code{DELEGATECALL}, the code is still threatened by the \code{SELFDESTRUCT} instruction in the executed code via \code{DELEGATECALL}.  
This issue is especially severe in EIP-1822 contracts, where callee contracts have upgrade capabilities instead of caller contracts. If the callee contract’s code is destroyed, the caller contract can no longer upgrade, rendering the entire upgradeable contract permanently unusable and leading to significant financial losses.

\noindent\textbf{Type II. Initialization of the caller contract (\ding{173}).}  \tab
Proper initialization of the caller contract is crucial for ensuring ownership security, which directly impacts fund security. Unlike traditional \code{Constructor}, which are automatically executed once during deployment, \code{initialize} functions are typically part of the callee contract and require explicit invocation from caller contracts. Developers must ensure these functions are executed correctly and only once, as failures can lead to ownership being compromised. For example, L2DAO lost 49.95M L2DAO tokens due to a failed initialization of the caller contract, allowing attackers to take over ownership. Similarly, Parity lost \$34M because its \code{initialize()} function was public and could be called multiple times, resetting ownership and enabling attackers to seize control. 

\noindent\textbf{Type III. Initialization of variables (\ding{174}).}  \tab
The Ronin Bridge incident, which resulted in a loss of \$12 million, was caused by insufficient initialization of variables introduced in an upgrade, resulting in a variable retaining an uninitialized value of 0 and being exploited by attackers to bypass security checks.  
For each new version, if new variables are introduced, there should be a function, such as \code{initializeV2()}, to initialize the new variables and ensure that they function correctly within the business logic.

\subsection{Collision}
\label{subsec:collision}

Collisions can be categorized into three types: storage collision, interface collision, and function selector collision.  
The first two types have resulted in actual security incidents, whereas the last type is discussed in relevant literature~\cite{bodell2023proxy} and community sources~\cite{function_selector_collision}, but no related incidents have been reported.  
For the sake of completeness in the insecurity taxonomy, we also introduce this type.


\noindent\textbf{Type I. Storage collision.}\tab
Storage collisions occur when the same storage slot is used inconsistently across contracts, which are in two scenarios: (1) between caller and callee contracts, and (2) among different versions of callee contracts.

In the first scenario, collisions between caller and callee contracts can overwrite critical variables. For example, Telcon, EmberSword, and DeltaPrime lost a total of approximately \$2.44M due to such collisions. In EmberSword, the caller contract used slot 0 to store the owner, while the callee contract used the same slot to store initialization flags. When the callee contract set the flags to prevent reinitialization, the caller contract overwrote slot 0, allowing attackers to bypass reinitialization checks and take over ownership.

In the second scenario, collisions between different versions of callee contracts can lead to outdated or incorrect values being used. EFVault, PikeFinance, and OKX NFT Aggregator lost a total of approximately \$6.9M due to this issue. For instance, in EFVault, a slot initially stored the maximum deposit limit (set to a very large value) in the first version. In the second version, the same slot was repurposed to store the asset decimal, a multiplier in valuation calculations. However, the slot’s content was not updated, causing the valuation logic to rely on the original large value, resulting in excessively high profit calculations and significant financial losses.

\noindent\textbf{Type II. Interface collision.}\tab
Interface collisions occur when the ABIs (Application Binary Interface) of an upgradeable contract are removed during an upgrade, disrupting interactions with other contracts, particularly non-upgradeable ones. For example, if the function \code{withdraw(uint256, uint256)} in the previous version of a callee contract is modified to \code{withdraw(uint256)} in the next version, the ABI for \code{withdraw(uint256, uint256)} is removed. This causes other contracts relying on the original ABI to fail, breaking compatibility.
When an upgradeable contract serves as a module within a DApp, the removal of ABIs can disrupt interactions, compromising the DApp’s overall availability. For instance, about \$110 million in AAVE were frozen for a week due to such an interface collision, underscoring the severe consequences of improper ABI management. The proper approach is to retain existing ABIs while adding new interfaces, ensuring compatibility between versions and preventing disruptions.



\noindent\textbf{Type III. Function selector collision.}\tab
Function selector collisions occur when the caller and callee contracts have functions with identical selectors. In cases where the caller forwards user input to the callee for execution, if a user inputs a function selector that also exists in the caller, the caller executes its own function instead of forwarding the call.
For example, if the caller contract has a function \code{collate\_propagate\_storage(bytes16)} and the callee contract has a function \code{burn(uint256)}, both sharing the selector \code{0x42966c68}, a user attempting to invoke \code{burn} via the caller contract would instead trigger \code{collate\_propagate\_storage}. If the caller’s function contains malicious code, such as unauthorized transfers, users may suffer financial losses.

\subsection{Flawed Business Logic}
\label{subsec:logic}

This subclass includes 12 security incidents resulting from flawed business logic introduced during upgrades, either through the addition of new functions or modifications to existing ones. These incidents encompass a range of vulnerabilities, such as public minting and accounting errors. While these vulnerabilities lack common semantics, they provide critical insights into security considerations during upgrade process.

\begin{figure}[t]
\centering
\footnotesize
\begin{lstlisting}[language=Solidity]
modifier onlyWhitelistMint() {
  require(!whitelistMint[msg.sender], "Invalid");_;}
function mint(...) public onlyWhitelistMint {...}
\end{lstlisting}
\vspace{-0.3cm}
\caption{The added public mint vulnerability in SafeMoon.}
\label{lst:safemoon_public_mint}
\end{figure}

\noindent\textbf{Insight I. Simple errors in new functions.}\tab
Some vulnerabilities arise from straightforward errors when introducing new functions. A notable example is the SafeMoon incident, where a mint function was added with an access control error, leading to a public mint vulnerability. As illustrated in Fig.~\ref{lst:safemoon_public_mint}\footnote{The comparison of two versions is available at \url{https://upgradehub.xyz/diffs/bscscan/0x42981d0bfbaf196529376ee702f2a9eb9092fcb5?selected=20}}, the access control requirement in Line 2 should be \code{whitelistMint[msg.sender]}, indicating that \code{msg.sender} must possess minting privileges. However, it was erroneously written as \code{!whitelistMint[msg.sender]}, allowing anyone except the privileged users to mint. Similar simple errors were observed in Level Finance and DEI.


\begin{figure}[t]
\centering
\begin{lstlisting}[language=Solidity]
function listToken(baseToken, correspondingToken, ...) {
`\colorbox{green!20}{+ }`if(_baseToken == _correspondingToken) {
`\colorbox{green!20}{+ }`  _tokenOwner[_baseToken] = msg.sender;} ...}
function withdrawTokens(token, ..., amount) {
`\colorbox{red!20}{- }`require(amount <= tokenDeposited[token][user],
`\colorbox{red!20}{- }`  "NOTHING_TO_CLAIM");
`\colorbox{green!20}{+ }`require(user == _tokenOwner[token],
`\colorbox{green!20}{+ }`  "ONLY_TOKEN_LISTER_CAN_WITHDRAW");}
\end{lstlisting}
\vspace{-0.3cm}
\caption{The introduced vulnerability in Xbridge.}
\label{lst:Xbridge}
\end{figure}

\noindent\textbf{Insight II. Risks in code modifications.}\tab
Modifying code, particularly when involving complex changes to data structures and multiple upgrades, can be inherently risky and may gradually introduce vulnerabilities. For instance, in Xbridge, a vulnerability was introduced over two upgrades, as depicted in Fig.~\ref{lst:Xbridge}. The vulnerability comprises two parts: Lines 2 and 3 in the \code{listToken} function enable attackers to designate themselves as token owners, and Lines 7 and 8 in the \code{withdrawTokens} function permit token owners to withdraw tokens without making a deposit. The first upgrade altered the code from Lines 5 to 7, while the second upgrade added Lines 2 and 3. Notably, both upgrades involved complex changes, such as introducing new ledgers and discarding old ones~\footnote{Here is the complete code, \url{https://upgradehub.xyz/diffs/etherscan/0x47ddb6a433b76117a98fbeab5320d8b67d468e31}}.

\begin{figure}[t]
\centering
\begin{lstlisting}[language=Solidity]
`\colorbox{red!20}{- }`function mint(...) external onlyMinter {
`\colorbox{green!20}{+ }`function mint(...) external {
\end{lstlisting}
\vspace{-0.3cm}
\caption{The introduced public mint vulnerability in Ankr.}
\label{lst:ANKR}
\end{figure}

\noindent\textbf{Insight III. Suspicious cases indicating potential malicious intent.}\tab
Certain cases appear suspicious and may indicate deliberate attacks or rug pulls. For example, in Ankr, the access control in the \code{mint} function was removed during an upgrade, introducing a public mint vulnerability. Immediately following the upgrade, the \code{mint} function was exploited for financial gain.

\subsection{Malicious Code Injection}
\label{subsec:malicious}

Malicious code injection accounts for 14 incidents, the highest among the four categories. Unlike the diversity observed in the flawed business logic category (\S\ref{subsec:logic}), these incidents exhibit significant consistency. We analyze the characteristics of malicious upgrades from three perspectives: root cause, introduction method, and code features.

\noindent\textbf{Feature I. Root cause.}\tab
Most malicious upgrades result from the theft of private keys. Two types of accounts are typically compromised: privileged EOAs of victim contracts and privileged EOAs of contracts governing multiple victim contracts. The latter is more severe, as it affects multiple contracts. For example, in Wilder World, the compromise of a governance contract enabled malicious upgrades across eight upgradeable contracts. Additionally, two incidents involved simultaneous compromises of multiple private keys, revealing poor private key management. In Opsec, both the privileged contract and the owner were compromised, while in Radiant, multi-signature measures failed due to multiple private key compromises. These incidents underscore the critical importance of private key security, as private keys grant the ability to modify code and access all assets of upgradeable contracts.

\begin{figure}[t]
\centering
\begin{lstlisting}[language=Solidity]
function withdraw( address token, address user, uint256 amount) external onlyOwner {
`\colorbox{red!20}{- }`if (token == address(opsec)) {
`\colorbox{red!20}{- }`  require(IERC20(token).balanceOf(address(this)) - totalStaked >= amount, "Insufficient $OPSEC balance to withdraw");}
   IERC20(token).safeTransfer(user, amount);}
\end{lstlisting}
\vspace{-0.3cm}
\caption{The introduced malicious code by changing code in Opsec.}
\label{lst:Opsec}
\end{figure}

\noindent\textbf{Feature II. Introduce method.}\tab
Malicious code is introduced in three ways: (1) replacing all original code with malicious code, (2) adding malicious code while retaining the original code, and (3) modifying the original code to make it malicious. Total replacement is a dangerous signal, as it disrupts the contract’s functionality. Adding malicious code while retaining the original code allows attackers to operate covertly, potentially yielding long-term gains, though it incurs higher attack costs. Modifying the original code requires it to resemble exploitative backdoor functions. For example, Fig.~\ref{lst:Opsec} shows malicious code introduced by removing checks in Lines 2 and 3, enabling the owner to arbitrarily obtain tokens.


\begin{figure}[t]
\centering
\footnotesize
\begin{lstlisting}[language=Solidity]
// Restrictions on executors 
require(msg.sender == owner, 'Ownable: caller isnot the owner'); // Meerkat Finance
require(msg.sender == 0x45..ec); // OCASH
address(0xcb..25) == tx.origin // Perpy Finance
// Restrictions on beneficiaries and profit mechanisms
address(0x36..c9).call().value(this.balance); // Polyhed
msg.sender.call().value(this.balance); // Shido
token.transferFrom(users, to, token.allowance(users,this)); // Atlantis Loans
token.transfer(msg.sender, token.balanceOf(address(this))) // Atlantis Loans
\end{lstlisting}
\vspace{-0.3cm}
\caption{Common malicious code.}
\label{lst:malicious}
\end{figure}

\noindent\textbf{Feature III. Code features.}\tab
Malicious code is typically simple and direct, designed to achieve its goals with minimal complexity. It generally features two main components: constraints and profit mechanisms. Constraints ensure that only the attacker benefits from the malicious code, often by restricting who can execute the code or who receives the profits. Executors are typically verified using \code{msg.sender} or \code{tx.origin}, with hardcoded addresses frequently used for validation. This approach contrasts with benign access controls, which often rely on complex data structures like storage mappings. Beneficiaries, on the other hand, are restricted to specific profit addresses, which are often hardcoded or limited to \code{msg.sender}. These constraints ensure that the attacker retains full control over the exploitation process.

Profit mechanisms in malicious code can be categorized into two types: indirect and direct. Indirect profiting occurs when attackers exploit privileged contracts to invoke functions like \code{mint} or \code{claimToken} on other contracts.
Direct profiting, on the other hand, targets the victim contract’s own assets, which can be categorized into three types: native tokens, valuable tokens (e.g., ERC-20, ERC-721), and approvals. Attackers transfer native tokens using \code{call.value}, valuable tokens via \code{token.transfer}, and approved assets via \code{token.transferFrom}. Profit amounts typically target the entire balance of the victim contract, such as \code{this.balance} for native tokens, \code{token.balanceOf} for valuable tokens, or \code{token.allowance} for approved assets.
This approach reflects a tendency to maximize gains by draining all available resources in a single attack.

\begin{framed}
    \vspace{-0.1in}
        \noindent
        \textbf{Answer to RQ2}\tab
            There are 37 real-world incidents involving over \$400 million in losses due to upgrades, with root causes categorized into eight distinct types.
        \vspace{-0.1in}
\end{framed}

\section{Evaluation of Upgrade Risks in the Wild (RQ3)}
\label{sec:RQ3}

Based on the taxonomy proposed in RQ2, we evaluate the insecurity of upgrades in the wild.  
For each type of security risk, we first conduct a survey (\S\ref{subsec:suvery}) to assess their current state of understanding and mitigation efforts.  
This survey aims to guide the community in understanding the current state of the art and identifying unresolved issues.
For risks with unknown impacts, we further utilize the dataset constructed in RQ1 to conduct large-scale evaluations (\S\ref{subsec:discovery}) to provide preliminary insights into these risks.


\subsection{Survey Results}
\label{subsec:suvery}
\begin{table}
\centering
\caption{Survey result about public awareness, scope of impact, and existing mitigation measures related to the insecurity identified in RQ2.}
\vspace{-0.3cm}
\label{tab:survey}
\footnotesize
\resizebox{\linewidth}{!}{%
\begin{tabular}{clccl}
\toprule
Category & \multicolumn{1}{c}{Subclass} & Public & Impact & Existing Mitigation \\ 
\midrule
\multirow{3}{*}{\begin{tabular}[c]{@{}c@{}}Improper\\Initialization\end{tabular}} & Callee & \yes & Limited\cite{li2024characterizing} & Auxiliary plugin and detectors \\ 
\cline{2-5}
 & Caller & \yes & Unknown & Auxiliary plugin \\ 
\cline{2-5}
 & Variables & \no & Unknown & No \\ 
\hline
\multirow{4}{*}{Collision} & Storage & \yes & Severe\cite{ruaro_not_2024} & \begin{tabular}[c]{@{}l@{}}EIP-1967\cite{EIP-1967}, EIP-7201\cite{EIP-7201}, and\\storage gaps~\cite{storage_gaps} \end{tabular} \\ 
\cline{2-5}
 & Interface & \no & Unknown & No \\ 
\cline{2-5}
 & \begin{tabular}[c]{@{}l@{}}Function\\Selector\end{tabular} & \yes & No\cite{bodell2023proxy} & No \\ 
\hline
\multicolumn{2}{c}{Flawed Business Logic} & \no & Unknown & No \\ 
\hline
\multicolumn{2}{c}{Malicious Code Injection} & \no & Unknown & No \\
\bottomrule
\end{tabular}
}
\end{table}

Our survey evaluates the public awareness, scope of impact, and existing mitigations for the insecurities we identified. To assess public awareness, we examine whether these vulnerabilities are documented in authoritative sources such as developer guides, well-known security blogs, and prior academic studies. For insecurities that have been studied, we derive their scope of impact from the findings of these studies. Additionally, we identify existing mitigations from community-provided security tools and proposed standards for upgradeable contracts.

The survey results are shown in Table~\ref{tab:survey}. Among the eight types of security risks in upgrades, five have unknown impacts, while four are entirely overlooked and lack any mitigation measures. Of the four risks known to the community, function selector collision is uncommon in practice, with no reported issues or incidents, and thus no mitigations exist. Initialization of the caller contract has limited discussion in the community, and no studies on its impact have been conducted; we address this in the following section (\S\ref{subsec:discovery}). In contrast, initialization of callee contracts and storage collisions are widely recognized in studies, with established mitigation measures.

Issues caused by uninitialized callee contracts have been extensively discussed in the community. Security vendors like OpenZeppelin (OZ) provide tools~\cite{oz_support} to mitigate this risk, including detectors that verify the security of callee contracts before deployment and plugins that enforce initialization during deployment. As a result, previous work~\cite{li2024characterizing} has studied its impact, finding limited occurrences, indicating effective management in practice.

Storage collision is another well-known issue in both the community and academic research.  
Several studies~\cite{bodell2023proxy,salehi2022not} have discussed this problem, and one study~\cite{ruaro_not_2024} proposed CRUSH, a detection tool that revealed a significant number of exploitable contracts due to storage collision flaws. To mitigate these collisions, the community has proposed methods such as standard slots (EIP-1967~\cite{EIP-1967}), storage gaps~\cite{storage_gaps}, and namespaced storage layouts (EIP-7201~\cite{EIP-7201}). EIP-1967 uses hashed keys for designated slots (e.g., admin and callee addresses) to minimize collision risks. Storage gaps and EIP-7201 partition storage to prevent layout errors between versions: storage gaps reserve intervals using fixed-size \code{uint256} arrays, while EIP-7201 organizes storage with structs and complex keys. Both methods require that variables not be removed and that new variables be added at the end within a single base contract.

\subsection{Discovered Issues}
\label{subsec:discovery}

\begin{table}
\centering
\caption{Discoveried Issues}
\label{tab:discovery}
\vspace{-0.3cm}
\footnotesize
\resizebox{\linewidth}{!}{%
\begin{tabular}{lcccccc} 
\toprule
 & \multicolumn{2}{c}{\begin{tabular}[c]{@{}c@{}}Improper\\Initialization\end{tabular}} & \multicolumn{2}{c}{\begin{tabular}[c]{@{}c@{}}Interface\\Collision\end{tabular}} & \multicolumn{2}{c}{\begin{tabular}[c]{@{}c@{}}Malicious Code\\Injection\end{tabular}} \\ 
\cmidrule(lr){2-3}\cmidrule(lr){4-5}\cmidrule(lr){6-7}
 & \begin{tabular}[c]{@{}c@{}}No\\Admin\end{tabular} & \begin{tabular}[c]{@{}c@{}}Delayed\\Init\end{tabular} & \begin{tabular}[c]{@{}c@{}}ABI\\Removal\end{tabular} & \begin{tabular}[c]{@{}c@{}}Output\\Change\end{tabular} & \begin{tabular}[c]{@{}c@{}}Authorized\\Transfer\end{tabular} & \begin{tabular}[c]{@{}c@{}}Authorized\\Set\end{tabular} \\ 
\hline
Ethereum & 10,374 & 80 & 1,990 & 284 & 1,556 & 1,884 \\
BSC & 9,245 & 107 & 1,933 & 333 & 1,509 & 2,112 \\ 
\hline
Total & 19,619 & 187 & 3,923 & 617 & 3,065 & 3,996 \\
\bottomrule
\end{tabular}
}
\end{table}

Among the five types of upgrade security risks with unknown impacts, we evaluate their security implications using the smart contract upgrade dataset from RQ1. Our findings reveal significant issues associated with three of these types. For the remaining two—flawed business logic and improper initialization of variables—evaluation requires automated understanding of code and variable semantics, which remains an open research challenge. Table~\ref{tab:discovery} summarizes the results.

\noindent\textbf{Initialization of caller contracts.}\tab
Previous studies on upgradeable contracts have overlooked this issue. To address this gap, we evaluate the initialization practices of caller contracts.

We focus on the time intervals between contract deployment and initialization, as a secure initialization process should ensure immediate ownership control by the project. Specifically, we record the deployment transaction as the deployment time and the first transaction setting the owner as the initialization time, using block intervals between them as the metric. The initialization transaction is identified by monitoring storage changes, leveraging EIP-1967, which specifies the slot key for storing the \code{admin} address.

We evaluate 42,057 upgradeable caller contracts following EIP-1967 in our dataset. Table~\ref{tab:discovery} presents the results. Notably, 19,619 contracts do not set the admin slot, and 187 contracts fail to initialize promptly after deployment. The high proportion of contracts neglecting the admin slot—nearly half of the total—is concerning. This often occurs because developers conflate the roles of the admin (responsible for upgrades) and the owner (responsible for business operations), directly using the owner for upgrades instead of the admin slot. This ambiguity in permission management poses significant risks, as the admin should only handle upgrades and avoid interacting with the callee contract.

For contracts that delay initialization, ownership is at risk during the vulnerable interval before initialization. Attackers have ample time to exploit this window, with observed intervals reaching 4,210 blocks on Ethereum and 28,636 blocks on BSC. If an attacker sets the admin slot first, the deployer loses control, potentially rendering the contract unusable and leading to asset loss as the L2DAO incident.

\noindent\textbf{Interface collision.}\tab
To study the impact of this issue, we crawl the source code information for upgraded contracts from RQ1 using blockchain explorers~\cite{etherscan,bscscan}. From this data, we extract ABIs and their outputs—whose changes trigger interface collisions—for each version of the callee contract and compare them. Table~\ref{tab:discovery} presents the results.

We identify 1,990 ABI removals and 284 interface output changes on Ethereum, and 1,933 ABI removals and 333 interface output changes on BSC. These issues can severely disrupt the availability of contracts that interact with the upgraded contracts, particularly non-upgradeable ones, as they become permanently unusable if the ABIs are removed during upgrades. These findings reveal that this issue is overlooked by the community and prior studies, and developers still lack awareness of compatibility, not only for storage but also for interfaces. As interactions between smart contracts become increasingly common, the impact of this issue can be significant, especially for large projects that rely on numerous interacting contracts and favor upgradeable contracts for maintenance. Developers interacting with upgradeable contracts should address this issue to ensure seamless interoperability for their own contracts.

\noindent\textbf{Suspicious Code Injection.}\tab
To study the impact of this issue, we leverage the code features of malicious code identified in RQ2 to detect similar code introduced during upgrades, using the dataset from RQ1. From RQ2, we extract code features related to constraints on executors and beneficiaries, as well as profit mechanisms. Using these features, we perform bytecode analysis to identify suspicious code introduced in upgrades. Specifically, constraints on executors are detected by identifying comparisons of executors with constants or storage variables, as shown in Expression~\ref{exp:executor}. Constraints on beneficiaries are detected when profit statement targets are constants or storage variables, as shown in Expression~\ref{exp:beneficiary}. Profit statements are detected by matching the three patterns specified in RQ2, corresponding to the three types of assets, as shown in Expression~\ref{exp:profit}. If the code introduced during upgrades matches any constraint pattern and the profit pattern, we classify it as suspicious.

\begin{equation}
\label{exp:executor}
   \frac{EQ(\code{msg.sender}|\code{tx.orogin}, Constant|SLOAD)}{ExecutorConstraint}
\end{equation}
\vspace{-0.1cm}
\begin{equation}
\label{exp:beneficiary}
   \frac{ProfitTarget(Constant|SLOAD)}{BeneficiaryConstraint}
\end{equation}
\vspace{-0.1cm}
\begin{equation}
\label{exp:profit}
   \frac{call.value|token.transfer|token.transferFrom}{Profit}
\end{equation}

\begin{figure}[t]
\centering
\begin{lstlisting}[language=Solidity]
// Arbitrarily transfer
function 0x27941c5b(varg0,varg1,varg2){
  require(msg.sender == 0x40e...ad0);
  address(varg0).transfer(address(varg2), varg1);}
function Withdraw(token,amount,_wallet) onlyOwner {
  if (token == address(0)) {
    _wallet.transfer(amount);}
  else {IERC20Upgradeable(token).transfer(_wallet, amount);}}
function releaseAllETH(account) onlyOwner {
  uint256 amount = address(this).balance;
  (bool success, )=account.call{value: amount}("");}
// Authority operation
function addBlacklist(account, value) onlyOwner {
  _isBlacklisted[account] = value;}
\end{lstlisting}
\vspace{-0.3cm}
\caption{Examples for detected suspicious code.}
\label{lst:suspicious}
\end{figure}

The results are shown in Table~\ref{tab:discovery}. Fig.~\ref{lst:suspicious} presents examples of detected code\footnote{Addresses for examples: 0x97841dc43ed42346bdb31c88027f23c989c98797 on Ethereum, 0xc3e9932ed58ed4131b7bd8155ace71659e8b12c5 on BSC, 0x247bc8cbb1a10bfb3dd159100c312e7add7ad7a9 on Ethereum, and 0xe1e2e1a3a8a1520392be499fbc4726ab4f6317c6 on Ethereum.}. In total, we detect suspicious code enabling arbitrary transfers in 1,556 upgrades on Ethereum and 1,509 upgrades on BSC. The examples in Fig.~\ref{lst:suspicious} Lines 2-11 all allow the owner or a specific account to arbitrarily transfer assets from the upgraded contracts. While such code may serve legitimate purposes, users and security researchers should remain vigilant, as it grants privileged authority that could threaten users' financial security.

Additionally, we detect authority operations introduced in 1,884 upgrades on Ethereum and 2,112 upgrades on BSC. These operations, while lacking explicit profit statements, can still impact users' assets. For example, as shown in the last example in Fig.~\ref{lst:suspicious} Lines 13-14, the upgrade introduces code allowing the owner to blacklist any user, preventing them from receiving token transfers. Such code is classified as a backdoor function by prior work~\cite{ma2023pied}, as it can compromise the security of users' assets.

In summary, we identify a significant amount of suspicious code introduced during upgrades, posing a real-world threat to user assets. This finding highlights the broad implications of this insecurity in upgradeable contracts. However, awareness and mitigation measures for this issue remain lacking in both the community and academic research.

\begin{framed}
    \vspace{-0.1in}
        \noindent
        \textbf{Answer to RQ3}\tab
            Among the eight types of upgrade risks, five are overlooked with unknown impacts, and a total of 31,407 related issues are detected.
        \vspace{-0.1in}
\end{framed}

\section{Discussion}
\label{sec:discussion}

\noindent\textbf{Limitations.}\tab  
Our study has limitations in identifying upgradeable contracts and collecting security incidents. For upgradeable contract identification, challenges arise when callers use storage mappings to store callee addresses, with function selectors from calldata as keys to calculate the slot index for the \code{DELEGATECALL} target. In such cases, a caller may have multiple callees, but we cannot retrieve all slots in the storage mapping without knowing all keys. We can only access slots with definite indices.
For security incident collection, limitations stem from the potential omission of incidents, particularly those involving small amounts of funds. Although we rely on four well-known security databases, we cannot guarantee the capture of all incidents on blockchains. Additionally, we identify incidents related to upgradeable contracts by locating \code{DELEGATECALL} in attack transactions and manually analyzing victim contracts, which may result in some incidents being overlooked. Nevertheless, we have collected as many incidents as possible to avoid missing widespread events, ensuring our taxonomy is as comprehensive as possible. This taxonomy will continue to evolve as more security incidents are gathered in the future.

\section{Related Work}
\label{sec:relatedwork}

The security of upgradeable contracts has attracted significant research efforts. Meisami et al.~\cite{bodell2023proxy} propose a taxonomy for proxy-based upgradeable contracts and develop USCHUNT, a tool for detecting such contracts and their security issues. Unlike their focus on proxy patterns, our study examines upgrade behaviors—the evolution of executable code within a contract account. Li et al.~\cite{li2024characterizing} define six upgrade patterns and analyze their security issues using USCHUNT, treating all functional changes as upgrades. In contrast, we focus on code evolution and provide a broader perspective on upgrade risks. Salehi et al.~\cite{salehi2022not} classify upgradeable contracts by call types and study access control in \code{DELEGATECALL}-based contracts, while we develop a comprehensive taxonomy of insecurities based on real-world incidents. Huang et al.~\cite{huang2024sword} examine proxy-based and metamorphic contracts, focusing on upgrade motivations, whereas our work emphasizes upgrade risks. Two studies~\cite{frowis2022not,chen2020finding} detect metamorphic contracts using \code{CREATE2}, while our study focuses on \code{DELEGATECALL}-based upgrades. To mitigate insecurities, Antonino et al.~\cite{antonino2022specification} propose a formal framework for secure upgrades, and Ruaro et al.~\cite{ruaro_not_2024} introduce CRUSH, a tool for detecting storage collisions. Our work complements these efforts by uncovering new insecurities and providing a dataset for further research.

Beyond the scope of upgradeable contracts, numerous studies have focused on the security of smart contracts. Several works analyze bytecode for program analysis~\cite{mythril,luu_making_2016,mossberg2019manticore,rattle,brent2018vandal,contro2021ethersolve,grech2019gigahorse,grech2022elipmoc,zhou2018erays} and vulnerability detection~\cite{bose2022sailfish,torres2018osiris,grech2018madmax,qian2023demystifying,wang2024efficiently,liao2022smartdagger,wang2024efficiently}. Our identification of upgradeable contracts and detection of malicious code build on these existing technologies. In response to these studies, large-scale experiments~\cite{ren2021empirical,chaliasos2024smart,li2024static,sendner2024large} have revealed the limitations of current program analysis techniques in practical semantic understanding, preventing us from evaluating two types of insecurities. Additionally, several surveys and studies have explored attacks across various types of smart contracts~\cite{su2021evil,zhou2020ever,zhang2023demystifying,zhou2023sok}, while our work specifically focuses on security incidents related to upgradeable contracts.

\section{Conclusion}

In this work, we conduct the first large-scale study on the security about upgrade behaviors in smart contracts.
To do so, we build the first large-scale dataset about upgrade behaviors by analyzing both transactions and bytecode.
In addition, we develop a thorough taxonomy of insecurities in upgrade behaviors from 37 real-world security incidents that involve over \$400 million.
Based on the dataset, we evaluate the security of upgrade behaviors following the taxonomy.
In summary, we report 8 different types of security risks in upgrade behaviors and find a total of 31,407 issues.

\bibliographystyle{IEEEtran}
\bibliography{reference}

\begin{thebibliography}{10}
\providecommand{\url}[1]{#1}
\csname url@samestyle\endcsname
\providecommand{\newblock}{\relax}
\providecommand{\bibinfo}[2]{#2}
\providecommand{\BIBentrySTDinterwordspacing}{\spaceskip=0pt\relax}
\providecommand{\BIBentryALTinterwordstretchfactor}{4}
\providecommand{\BIBentryALTinterwordspacing}{\spaceskip=\fontdimen2\font plus
\BIBentryALTinterwordstretchfactor\fontdimen3\font minus \fontdimen4\font\relax}
\providecommand{\BIBforeignlanguage}[2]{{%
\expandafter\ifx\csname l@#1\endcsname\relax
\typeout{** WARNING: IEEEtran.bst: No hyphenation pattern has been}%
\typeout{** loaded for the language `#1'. Using the pattern for}%
\typeout{** the default language instead.}%
\else
\language=\csname l@#1\endcsname
\fi
#2}}
\providecommand{\BIBdecl}{\relax}
\BIBdecl

\bibitem{frowis2022not}
M.~Fr{\"o}wis and R.~B{\"o}hme, ``Not all code are create2 equal,'' in \emph{International Conference on Financial Cryptography and Data Security}.\hskip 1em plus 0.5em minus 0.4em\relax Springer, 2022, pp. 516--538.

\bibitem{chen2020finding}
J.~Chen, ``Finding ethereum smart contracts security issues by comparing history versions,'' in \emph{Proceedings of the 35th IEEE/ACM International Conference on Automated Software Engineering}, 2020, pp. 1382--1384.

\bibitem{huang2024sword}
Y.~Huang, X.~Wu, Q.~Wang, Z.~Qian, X.~Chen, M.~Tang, and Z.~Zheng, ``The sword of damocles: Upgradeable smart contract in ethereum,'' in \emph{Proceedings of the 32nd IEEE/ACM International Conference on Program Comprehension}, 2024, pp. 333--345.

\bibitem{li2024characterizing}
X.~Li, J.~Yang, J.~Chen, Y.~Tang, and X.~Gao, ``Characterizing ethereum upgradable smart contracts and their security implications,'' in \emph{Proceedings of the ACM on Web Conference 2024}, 2024, pp. 1847--1858.

\bibitem{salehi2022not}
M.~Salehi, J.~Clark, and M.~Mannan, ``Not so immutable: Upgradeability of smart contracts on ethereum,'' in \emph{International Conference on Financial Cryptography and Data Security}.\hskip 1em plus 0.5em minus 0.4em\relax Springer, 2022, pp. 539--554.

\bibitem{chaliasos2024smart}
S.~Chaliasos, M.~A. Charalambous, L.~Zhou, R.~Galanopoulou, A.~Gervais, D.~Mitropoulos, and B.~Livshits, ``Smart contract and defi security tools: Do they meet the needs of practitioners?'' in \emph{Proceedings of the 46th IEEE/ACM International Conference on Software Engineering}, 2024, pp. 1--13.

\bibitem{bodell2023proxy}
W.~E. Bodell~III, S.~Meisami, and Y.~Duan, ``Proxy hunting: understanding and characterizing proxy-based upgradeable smart contracts in blockchains,'' in \emph{32nd USENIX Security Symposium (USENIX Security 23)}, 2023, pp. 1829--1846.

\bibitem{proxy_dataset}
``Proxy dataset,'' 2025, \url{https://github.com/USCHunt-Anon/USCHunt/tree/master/study/data}.

\bibitem{wood2014ethereum}
G.~Wood \emph{et~al.}, ``Ethereum: A secure decentralised generalised transaction ledger,'' \emph{Ethereum project yellow paper}, vol. 151, no. 2014, pp. 1--32, 2014.

\bibitem{bscscan}
``Bscscan: Bnb smart chain explorer,'' 2024, \url{https://bscscan.com/}.

\bibitem{erc20}
``Erc-20: Token standard,'' 2015, \url{https://eips.ethereum.org/EIPS/eip-20}.

\bibitem{erc721}
``Erc-721: Non-fungible token standard,'' 2018, \url{https://eips.ethereum.org/EIPS/eip-721}.

\bibitem{etherscan}
``Etherscan: The ethereum blockchain explorer,'' 2024, \url{https://etherscan.io/}.

\bibitem{upgradehub}
``Upgradehub,'' 2024, \url{https://upgradehub.xyz/}.

\bibitem{blocksec}
``Blocksec security incidents,'' 2024, \url{https://app.blocksec.com/explorer/security-incidents}.

\bibitem{defiLlama}
``Defillama hacks,'' 2024, \url{https://defillama.com/hacks}.

\bibitem{de.fi}
``De.fi rekt-database,'' 2024, \url{https://de.fi/rekt-database}.

\bibitem{slowmist}
``Slowmist hacked,'' 2024, \url{https://hacked.slowmist.io/}.

\bibitem{function_selector_collision}
``Beware of the proxy: learn how to exploit function clashing,'' 2019, \url{https://forum.openzeppelin.com/t/beware-of-the-proxy-learn-how-to-exploit-function-clashing/1070}.

\bibitem{ruaro_not_2024}
\BIBentryALTinterwordspacing
N.~Ruaro, F.~Gritti, R.~McLaughlin, I.~Grishchenko, C.~Kruegel, and G.~Vigna, ``\BIBforeignlanguage{en}{Not your {Type}! {Detecting} {Storage} {Collision} {Vulnerabilities} in {Ethereum} {Smart} {Contracts}},'' in \emph{\BIBforeignlanguage{en}{Proceedings 2024 {Network} and {Distributed} {System} {Security} {Symposium}}}.\hskip 1em plus 0.5em minus 0.4em\relax San Diego, CA, USA: Internet Society, 2024. [Online]. Available: \url{https://www.ndss-symposium.org/wp-content/uploads/2024-713-paper.pdf}
\BIBentrySTDinterwordspacing

\bibitem{EIP-1967}
``Erc-1967: Proxy storage slots,'' 2019, \url{https://eips.ethereum.org/EIPS/eip-1967}.

\bibitem{EIP-7201}
``Erc-7201: Namespaced storage layout,'' 2023, \url{https://eips.ethereum.org/EIPS/eip-7201}.

\bibitem{storage_gaps}
``Storage gaps,'' 2025, \url{https://docs.openzeppelin.com/upgrades-plugins/writing-upgradeable#storage-gaps}.

\bibitem{oz_support}
``Openzeppelin upgradesn,'' 2025, \url{https://docs.openzeppelin.com/upgrades}.

\bibitem{ma2023pied}
F.~Ma, M.~Ren, L.~Ouyang, Y.~Chen, J.~Zhu, T.~Chen, Y.~Zheng, X.~Dai, Y.~Jiang, and J.~Sun, ``Pied-piper: Revealing the backdoor threats in ethereum erc token contracts,'' \emph{ACM Transactions on Software Engineering and Methodology}, vol.~32, no.~3, pp. 1--24, 2023.

\bibitem{antonino2022specification}
P.~Antonino, J.~Ferreira, A.~Sampaio, and A.~Roscoe, ``Specification is law: Safe creation and upgrade of ethereum smart contracts,'' in \emph{International Conference on Software Engineering and Formal Methods}.\hskip 1em plus 0.5em minus 0.4em\relax Springer, 2022, pp. 227--243.

\bibitem{mythril}
``Mythril,'' 2022, \url{https://github.com/ConsenSys/mythril}.

\bibitem{luu_making_2016}
L.~Luu, D.-H. Chu, H.~Olickel, P.~Saxena, and A.~Hobor, ``Making smart contracts smarter,'' in \emph{Proceedings of the 2016 ACM SIGSAC conference on computer and communications security}, 2016, pp. 254--269.

\bibitem{mossberg2019manticore}
M.~Mossberg, F.~Manzano, E.~Hennenfent, A.~Groce, G.~Grieco, J.~Feist, T.~Brunson, and A.~Dinaburg, ``Manticore: A user-friendly symbolic execution framework for binaries and smart contracts,'' in \emph{2019 34th IEEE/ACM International Conference on Automated Software Engineering (ASE)}.\hskip 1em plus 0.5em minus 0.4em\relax IEEE, 2019, pp. 1186--1189.

\bibitem{rattle}
``Rattle,'' 2022, \url{https://github.com/crytic/rattle}.

\bibitem{brent2018vandal}
L.~Brent, A.~Jurisevic, M.~Kong, E.~Liu, F.~Gauthier, V.~Gramoli, R.~Holz, and B.~Scholz, ``Vandal: A scalable security analysis framework for smart contracts,'' \emph{arXiv preprint arXiv:1809.03981}, 2018.

\bibitem{contro2021ethersolve}
F.~Contro, M.~Crosara, M.~Ceccato, and M.~Dalla~Preda, ``Ethersolve: Computing an accurate control-flow graph from ethereum bytecode,'' in \emph{2021 IEEE/ACM 29th International Conference on Program Comprehension (ICPC)}.\hskip 1em plus 0.5em minus 0.4em\relax IEEE, 2021, pp. 127--137.

\bibitem{grech2019gigahorse}
N.~Grech, L.~Brent, B.~Scholz, and Y.~Smaragdakis, ``Gigahorse: thorough, declarative decompilation of smart contracts,'' in \emph{2019 IEEE/ACM 41st International Conference on Software Engineering (ICSE)}.\hskip 1em plus 0.5em minus 0.4em\relax IEEE, 2019, pp. 1176--1186.

\bibitem{grech2022elipmoc}
N.~Grech, S.~Lagouvardos, I.~Tsatiris, and Y.~Smaragdakis, ``Elipmoc: Advanced decompilation of ethereum smart contracts,'' \emph{Proceedings of the ACM on Programming Languages}, vol.~6, no. OOPSLA1, pp. 1--27, 2022.

\bibitem{zhou2018erays}
Y.~Zhou, D.~Kumar, S.~Bakshi, J.~Mason, A.~Miller, and M.~Bailey, ``Erays: reverse engineering ethereum's opaque smart contracts,'' in \emph{27th USENIX security symposium (USENIX Security 18)}, 2018, pp. 1371--1385.

\bibitem{bose2022sailfish}
P.~Bose, D.~Das, Y.~Chen, Y.~Feng, C.~Kruegel, and G.~Vigna, ``Sailfish: Vetting smart contract state-inconsistency bugs in seconds,'' in \emph{2022 IEEE Symposium on Security and Privacy (SP)}.\hskip 1em plus 0.5em minus 0.4em\relax IEEE, 2022, pp. 161--178.

\bibitem{torres2018osiris}
C.~F. Torres, J.~Sch{\"u}tte, and R.~State, ``Osiris: Hunting for integer bugs in ethereum smart contracts,'' in \emph{Proceedings of the 34th annual computer security applications conference}, 2018, pp. 664--676.

\bibitem{grech2018madmax}
N.~Grech, M.~Kong, A.~Jurisevic, L.~Brent, B.~Scholz, and Y.~Smaragdakis, ``Madmax: Surviving out-of-gas conditions in ethereum smart contracts,'' \emph{Proceedings of the ACM on Programming Languages}, vol.~2, no. OOPSLA, pp. 1--27, 2018.

\bibitem{qian2023demystifying}
P.~Qian, J.~He, L.~Lu, S.~Wu, Z.~Lu, L.~Wu, Y.~Zhou, and Q.~He, ``Demystifying random number in ethereum smart contract: taxonomy, vulnerability identification, and attack detection,'' \emph{IEEE Transactions on Software Engineering}, vol.~49, no.~7, pp. 3793--3810, 2023.

\bibitem{wang2024efficiently}
Z.~Wang, J.~Chen, Y.~Wang, Y.~Zhang, W.~Zhang, and Z.~Zheng, ``Efficiently detecting reentrancy vulnerabilities in complex smart contracts,'' \emph{arXiv preprint arXiv:2403.11254}, 2024.

\bibitem{liao2022smartdagger}
Z.~Liao, Z.~Zheng, X.~Chen, and Y.~Nan, ``Smartdagger: a bytecode-based static analysis approach for detecting cross-contract vulnerability,'' in \emph{Proceedings of the 31st ACM SIGSOFT International Symposium on Software Testing and Analysis}, 2022, pp. 752--764.

\bibitem{ren2021empirical}
M.~Ren, Z.~Yin, F.~Ma, Z.~Xu, Y.~Jiang, C.~Sun, H.~Li, and Y.~Cai, ``Empirical evaluation of smart contract testing: What is the best choice?'' in \emph{Proceedings of the 30th ACM SIGSOFT international symposium on software testing and analysis}, 2021, pp. 566--579.

\bibitem{li2024static}
K.~Li, Y.~Xue, S.~Chen, H.~Liu, K.~Sun, M.~Hu, H.~Wang, Y.~Liu, and Y.~Chen, ``Static application security testing (sast) tools for smart contracts: How far are we?'' \emph{Proceedings of the ACM on Software Engineering}, vol.~1, no. FSE, pp. 1447--1470, 2024.

\bibitem{sendner2024large}
C.~Sendner, L.~Petzi, J.~Stang, and A.~Dmitrienko, ``Large-scale study of vulnerability scanners for ethereum smart contracts,'' in \emph{2024 IEEE Symposium on Security and Privacy (SP)}.\hskip 1em plus 0.5em minus 0.4em\relax IEEE Computer Society, 2024, pp. 220--220.

\bibitem{su2021evil}
L.~Su, X.~Shen, X.~Du, X.~Liao, X.~Wang, L.~Xing, and B.~Liu, ``Evil under the sun: Understanding and discovering attacks on ethereum decentralized applications,'' in \emph{30th USENIX Security Symposium (USENIX Security 21)}, 2021, pp. 1307--1324.

\bibitem{zhou2020ever}
S.~Zhou, M.~M{\"o}ser, Z.~Yang, B.~Adida, T.~Holz, J.~Xiang, S.~Goldfeder, Y.~Cao, M.~Plattner, X.~Qin \emph{et~al.}, ``An ever-evolving game: Evaluation of real-world attacks and defenses in ethereum ecosystem,'' in \emph{29th USENIX Security Symposium (USENIX Security 20)}, 2020, pp. 2793--2810.

\bibitem{zhang2023demystifying}
Z.~Zhang, B.~Zhang, W.~Xu, and Z.~Lin, ``Demystifying exploitable bugs in smart contracts,'' in \emph{2023 IEEE/ACM 45th International Conference on Software Engineering (ICSE)}.\hskip 1em plus 0.5em minus 0.4em\relax IEEE, 2023, pp. 615--627.

\bibitem{zhou2023sok}
L.~Zhou, X.~Xiong, J.~Ernstberger, S.~Chaliasos, Z.~Wang, Y.~Wang, K.~Qin, R.~Wattenhofer, D.~Song, and A.~Gervais, ``Sok: Decentralized finance (defi) attacks,'' in \emph{2023 IEEE Symposium on Security and Privacy (SP)}.\hskip 1em plus 0.5em minus 0.4em\relax IEEE, 2023, pp. 2444--2461.

\end{thebibliography}

\end{document}